\documentclass[a4paper,draft,oneside,notitlepage]{article}
\usepackage{sw20lart}

\input tcilatex
\QQQ{Language}{
British English
}

\begin{document}

\section{Introduction}

In previous papers, we have shown that the standard description of spin can
be generalized[1-7]. We have shown that more generalized probability
amplitudes than the standard forms in the literature can be derived for the
cases of spin 1/2, spin 1, and the singlet and triplet systems resulting
from the addition of the spins of two spin 1/2 systems. There is no reason
to suppose that we could not obtain similar generalized results for any
other spin system, though at this stage we have only been able to treat the
systems mentioned above.

The reasoning used in deriving the generalized results naturally leads us to
ask: in view of the similarities between intrinsic spin and orbital angular
momentum, is it possible to carry out a programme of generalization for the
ordinary spherical harmonics? If generalized spherical harmonics do indeed
exist, are they solutions of some differential eigenvalue equation? We show
in this paper that the answers to these questions are affirmative.

\section{General Theory}

\subsection{Probability Amplitudes}

Our considerations are based on the interpretation of quantum mechanics due
to Land\'e[8-11]. In this interpretation, a wave function is fundamentally a
probability amplitude connecting an initial and a final state of a system.
In consequence, it is characterized by two sets of quantum numbers - a set
corresponding to the initial state, and a set corresponding to the final
state. To give an example, the eigenfunctions $\psi _n(\mathbf{r})$ arising
from solution of the time-independent Schr\"odinger equation are probability
amplitudes connecting two states. For each eigenfunction, the initial state
corresponds to the energy eigenvalue $E_n$, and the final state to the
position $\mathbf{r}$. Similarly, the spherical harmonic $Y_m^l(\theta
,\varphi )$ is a probability amplitude referring to an initial state in
which the angular momentum projection is $m\hbar $ along the $z$ axis, while
the final state is defined by the angular position $(\theta ,\varphi ).$

For a spin system, we denote the probability amplitudes by $\chi (m_i^{(%
\widehat{\mathbf{a}})};m_f^{(\widehat{\mathbf{c}})}).$ The probability
amplitude $\chi (m_i^{(\widehat{\mathbf{a}})};m_f^{(\widehat{\mathbf{c}})})$
yields the probability that if the spin projection of the system in the
direction $\widehat{\mathbf{a}}$ (whose polar angles are $(\theta ^{\prime
},\varphi ^{\prime })$) is $m_i\hbar $, then a measurement of the spin
projection in another direction $\widehat{\mathbf{c}}$ (whose polar angles
are $(\theta ,\varphi )$) yields the projection $m_f\hbar ,$ where $m_i$ and 
$m_f$ are the projection quantum numbers. For any spin system, the
generalized probability amplitudes take many forms because different choices
of phase are possible[4]. For a spin-1 system, one form of these probability
amplitudes is [3]:

\begin{eqnarray}
\chi ((+1)^{(\widehat{\mathbf{a}})};(+1)^{(\widehat{\mathbf{c}})}) &=&\cos ^2%
\frac{\theta ^{\prime }}2\cos ^2\frac \theta 2e^{-i(\varphi ^{\prime
}-\varphi )}+\sin ^2\frac{\theta ^{\prime }}2\sin ^2\frac \theta
2e^{i(\varphi ^{\prime }-\varphi )}  \nonumber \\
&&+\frac 12\sin \theta ^{\prime }\sin \theta ,  \label{tw2}
\end{eqnarray}

\begin{eqnarray}
\chi ((+1)^{(\widehat{\mathbf{a}})};0^{(\widehat{\mathbf{c}})}) &=&\frac 1{%
\sqrt{2}}[\sin ^2\frac{\theta ^{\prime }}2\sin \theta e^{i(\varphi ^{\prime
}-\varphi )}-\cos ^2\frac{\theta ^{\prime }}2\sin \theta e^{-i(\varphi
^{\prime }-\varphi )}  \nonumber  \label{fo4} \\
&&+\sin \theta ^{\prime }\cos \theta ],  \label{th3}
\end{eqnarray}

\begin{eqnarray}
\chi ((+1)^{(\widehat{\mathbf{a}})};(-1)^{(\widehat{\mathbf{c}})}) &=&\cos ^2%
\frac{\theta ^{\prime }}2\sin ^2\frac \theta 2e^{-i(\varphi ^{\prime
}-\varphi )}+\sin ^2\frac{\theta ^{\prime }}2\cos ^2\frac \theta
2e^{i(\varphi ^{\prime }-\varphi )}  \nonumber \\
&&\ -\frac 12\sin \theta ^{\prime }\sin \theta ,  \label{fo4}
\end{eqnarray}
\begin{eqnarray}
\chi (0^{(\widehat{\mathbf{a}})};(+1)^{(\widehat{\mathbf{c}})}) &=&\frac 1{%
\sqrt{2}}[-\sin \theta ^{\prime }\cos ^2\frac \theta 2e^{-i(\varphi ^{\prime
}-\varphi )}+\sin \theta ^{\prime }\sin ^2\frac \theta 2e^{i(\varphi
^{\prime }-\varphi )}  \nonumber \\
&&\ +\cos \theta ^{\prime }\sin \theta ],  \label{fi5}
\end{eqnarray}
\begin{equation}
\chi (0^{(\widehat{\mathbf{a}})};0^{(\widehat{\mathbf{c}})})=\frac 12\sin
\theta ^{\prime }\sin \theta e^{-i(\varphi ^{\prime }-\varphi )}+\frac
12\sin \theta ^{\prime }\sin \theta e^{i(\varphi ^{\prime }-\varphi )}+\cos
\theta ^{\prime }\cos \theta ,  \label{si6}
\end{equation}
\begin{eqnarray}
\chi (0^{(\widehat{\mathbf{a}})};(-1)^{(\widehat{\mathbf{c}})}) &=&\frac 1{%
\sqrt{2}}[-\sin \theta ^{\prime }\sin ^2\frac \theta 2e^{-i(\varphi ^{\prime
}-\varphi )}+\sin \theta ^{\prime }\cos ^2\frac \theta 2e^{i(\varphi
^{\prime }-\varphi )}  \nonumber \\
&&\ -\cos \theta ^{\prime }\sin \theta ],  \label{se7}
\end{eqnarray}
\begin{eqnarray}
\chi ((-1)^{(\widehat{\mathbf{a}})};(+1)^{(\widehat{\mathbf{c}})}) &=&\sin ^2%
\frac{\theta ^{\prime }}2\cos ^2\frac \theta 2e^{-i(\varphi ^{\prime
}-\varphi )}+\cos ^2\frac{\theta ^{\prime }}2\sin ^2\frac \theta
2e^{i(\varphi ^{\prime }-\varphi )}  \nonumber \\
&&\ -\frac 12\sin \theta ^{\prime }\sin \theta ,  \label{ei8}
\end{eqnarray}
\begin{eqnarray}
\chi ((-1)^{(\widehat{\mathbf{a}})};0^{(\widehat{\mathbf{c}})}) &=&\frac 1{%
\sqrt{2}}[-\sin ^2\frac{\theta ^{\prime }}2\sin \theta e^{-i(\varphi
^{\prime }-\varphi )}+\cos ^2\frac{\theta ^{\prime }}2\sin \theta
e^{i(\varphi ^{\prime }-\varphi )}  \nonumber \\
&&\ -\sin \theta ^{\prime }\cos \theta ]  \label{ni9}
\end{eqnarray}
and

\begin{eqnarray}
\chi ((-1)^{(\widehat{\mathbf{a}})};(-1)^{(\widehat{\mathbf{c}})}) &=&\sin ^2%
\frac{\theta ^{\prime }}2\sin ^2\frac \theta 2e^{-i(\varphi ^{\prime
}-\varphi )}+\cos ^2\frac{\theta ^{\prime }}2\cos ^2\frac \theta
2e^{i(\varphi ^{\prime }-\varphi )}  \nonumber \\
&&\ \ +\frac 12\sin \theta ^{\prime }\sin \theta ].  \label{te10}
\end{eqnarray}
These probability amplitudes reduce to the standard forms if $\theta
=\varphi =0,$ which happens if the direction $\widehat{\mathbf{c}}$ is taken
to be the $z$ direction[3].

\subsection{Law of Probability Addition}

Let $A$, $B$ and $C$ be observables of a quantum system with respective
eigenvalue spectra $A_1$, $A_2,...$, $B_1$, $B_2,...$, and $C_1$,$C_2,...$,
respectively. Then there are three sets of probability amplitudes belonging
to this system. Let $\psi (A_i,C_n)$ be the probability amplitudes for
measurements of $C$ when the system is in an eigenstate $A$. Let $\xi
(A_i,B_j)$ be the probability amplitudes for measurements of $B$ when the
system is in an eigenstate $A.$ Let $\phi (B_j,C_n)$ be the probability
amplitudes for measurements of $C$ when the system is in an eigenstate of $%
B. $ Then the three sets of probability amplitudes are connected by the law

\begin{equation}
\psi (A_i,C_n)=\sum_j\xi (A_i,B_j)\phi (B_j,C_n).  \label{el11}
\end{equation}
The probability amplitudes obey a Hermiticity condition because the
corresponding probabilities $P$ are symmetrical with regard to the
measurement direction[8]. Thus, because $P(A_i,C_n)=P(C_n,A_i)$ for example,
the condition

\begin{equation}
\psi (A_i,C_n)=\psi ^{*}(C_n,A_i)  \label{tw12}
\end{equation}
is satisfied.

\section{Application to Orbital Angular Momentum}

\subsection{Spherical Harmonics}

With these preliminaries, we can consider the spherical harmonics. As
observed above, they are fundamentally probability amplitudes. The spherical
harmonic $Y_m^l(\theta ,\varphi )$ is the probability amplitude that if the
system is in the state of angular momentum projection $m\hbar $ along the $z$
direction, then a measurement of the angular position of the system gives
the value $(\theta ,\varphi )$ in the solid element $d\Omega $[6].

It is evident that the spherical harmonics are generalizable. If the initial
quantization direction is not the $z$ direction but is defined by the unit
vector $\widehat{\mathbf{a}},$ then the probability amplitude corresponding
to measurement of the angular position becomes a generalized spherical
harmonic.

Let us denote the generalized spherical harmonics by $Y(l,m^{(\widehat{%
\mathbf{a}})};\theta ,\varphi )$. In this notation, the ordinary spherical
harmonics are denoted by $Y(l,m^{(\widehat{\mathbf{k}})};\theta ,\varphi ),$
since $\widehat{\mathbf{k}}$ is the unit vector defining the $z$ direction.
The generalized spherical harmonics can be constructed with the aid of Eq. (%
\ref{el11}). In order to do this, we make the identification $\psi
(A_i,C_n)=Y(l,m_i^{(\widehat{\mathbf{a}})};\theta ,\varphi ),$ so that $A$
is the angular momentum projection in the direction $\widehat{\mathbf{a}}$,
while $C$ is the angular position of the system. We also need an appropriate
third observable $B$ to facilitate the expansion Eq. (\ref{el11}). The
choice of $B$ must result in the probability amplitudes $\xi (A_i,B_j)$ and $%
\phi (B_j,C_n)$ being known or easy to deduce. We choose $B$ to be the
angular momentum projection in the $z$ direction. This choice indeed results
in the probability amplitudes $\xi (A_i,B_j)$ and $\phi (B_j,C_n)$ in the
expansion being known.

Since $A$ is an angular momentum projection and $B$ is an angular momentum
projection, the $\xi (A_i,B_j)$ are probability amplitudes for angular
momentum projection measurements from one direction to another. The initial
projection direction is $\widehat{\mathbf{a}}$, while the final projection
direction is the $z$ direction. For a given value of $l$ the probability
amplitudes for angular momentum projection measurements from one direction
to another should be identical to the probability amplitudes $\chi (m_i^{(%
\widehat{\mathbf{a}})};m_f^{(\widehat{\mathbf{c}})})$ for spin projection
measurements for these directions for $s=l.$

Since $B$ corresponds to angular momentum projection along the $z$ direction
and $C$ is the angular position of the system, it follows that the $\phi
(B_j,C_n)$ are just the ordinary spherical harmonics. Thus, the expansion
for the generalized spherical harmonics becomes

\begin{equation}
Y(l,m_i^{(\widehat{\mathbf{a}})};\theta ,\varphi )=\sum_j\chi (l,m_i^{(%
\widehat{\mathbf{a}})};l,m_j^{(\widehat{\mathbf{k}})})Y(l,m_j^{(\widehat{%
\mathbf{k}})};\theta ,\varphi ).  \label{th13}
\end{equation}

\subsection{Generalized Spherical Harmonics for $l=1$}

The case $l=1$ is the simplest to which we can apply these arguments. Then
the probability amplitudes $\chi (l,m_i^{(\widehat{\mathbf{a}})};l,m_j^{(%
\widehat{\mathbf{k}})})$ are specialized forms of the expressions for $s=1$.
They result if we set $\widehat{\mathbf{c}}=\widehat{\mathbf{k}}$ so that $%
\theta =\varphi =0$ in Eqs. (\ref{tw2})-(\ref{te10}). We obtain

\begin{eqnarray}
\chi ((+1)^{(\widehat{\mathbf{a}})};(+1)^{(\widehat{\mathbf{k}})})=\cos ^2%
\frac{\theta ^{\prime }}2e^{-i\varphi ^{\prime }},  \label{fo14}
\end{eqnarray}

\begin{equation}
\chi ((+1)^{(\widehat{\mathbf{a}})};0^{(\widehat{\mathbf{k}})})=\frac 1{%
\sqrt{2}}\sin \theta ^{\prime },  \label{fi15}
\end{equation}

\begin{equation}
\chi ((+1)^{(\widehat{\mathbf{a}})};(-1)^{(\widehat{\mathbf{k}})})=\sin ^2%
\frac{\theta ^{\prime }}2e^{i\varphi ^{\prime }},  \label{si16}
\end{equation}
\begin{equation}
\chi (0^{(\widehat{\mathbf{a}})};(+1)^{(\widehat{\mathbf{k}})})=-\frac 1{%
\sqrt{2}}\sin \theta ^{\prime }e^{-i\varphi ^{\prime }},  \label{se17}
\end{equation}
\begin{equation}
\chi (0^{(\widehat{\mathbf{a}})};0^{(\widehat{\mathbf{k}})})=\cos \theta
^{\prime },  \label{ei18}
\end{equation}
\begin{equation}
\chi (0^{(\widehat{\mathbf{a}})};(-1)^{(\widehat{\mathbf{k}})})=\frac 1{%
\sqrt{2}}\sin \theta ^{\prime }e^{i\varphi ^{\prime }},  \label{ni19}
\end{equation}
\begin{equation}
\chi ((-1)^{(\widehat{\mathbf{a}})};(+1)^{(\widehat{\mathbf{k}})})=-\sin ^2%
\frac{\theta ^{\prime }}2e^{-i\varphi ^{\prime }},  \label{tw20}
\end{equation}
\begin{equation}
\chi ((-1)^{(\widehat{\mathbf{a}})};0^{(\widehat{\mathbf{k}})})=\frac 1{%
\sqrt{2}}\sin \theta ^{\prime }  \label{tw21}
\end{equation}
and

\begin{equation}
\chi ((-1)^{(\widehat{\mathbf{a}})};(-1)^{(\widehat{\mathbf{k}})})=-\cos ^2%
\frac{\theta ^{\prime }}2e^{i\varphi ^{\prime }}.  \label{tw22}
\end{equation}

The ordinary spherical harmonics $Y_m^l(\theta ,\varphi )=Y(l,m^{(\widehat{%
\mathbf{k}})};\theta ,\varphi )$ for $l=1$ are 
\begin{equation}
Y(1,1^{(\widehat{\mathbf{k}})};\theta ,\varphi )=-\sqrt{\frac 3{8\pi }}\sin
\theta e^{i\varphi },  \label{tw23}
\end{equation}

\begin{equation}
Y(1,0^{(\widehat{\mathbf{k}})};\theta ,\varphi )=\sqrt{\frac 3{4\pi }}\cos
\theta  \label{tw24}
\end{equation}
and

\begin{equation}
Y(1,(-1)^{(\widehat{\mathbf{k}})};\theta ,\varphi )=\sqrt{\frac 3{8\pi }}%
\sin \theta e^{-i\varphi }.  \label{tw25}
\end{equation}

The summation over $j$ in Eq. (\ref{th13}) produces the values $m_j^{(%
\widehat{\mathbf{k}})}=-1,0,1$. Thus, the generalized spherical harmonics are

\begin{equation}
Y(1,1^{(\widehat{\mathbf{a}})};\theta ,\varphi )=\sqrt{\frac 3{8\pi }}[\cos
\theta ^{\prime }\sin \theta \cos (\varphi ^{\prime }-\varphi )+\sin \theta
^{\prime }\cos \theta +i\sin \theta \sin (\varphi ^{\prime }-\varphi )],
\label{tw26}
\end{equation}

\begin{equation}
Y(1,0^{(\widehat{\mathbf{a}})};\theta ,\varphi )=\sqrt{\frac 3{4\pi }}[\cos
\theta \cos \theta ^{\prime }+\sin \theta \sin \theta ^{\prime }\cos
(\varphi ^{\prime }-\varphi )]  \label{tw27}
\end{equation}
and

\begin{equation}
Y(1,(-1)^{(\widehat{\mathbf{a}})};\theta ,\varphi )=\sqrt{\frac 3{8\pi }}[%
-\sin \theta \cos \theta ^{\prime }\cos (\varphi ^{\prime }-\varphi )+\sin
\theta ^{\prime }\cos \theta -i\sin \theta \sin (\varphi ^{\prime }-\varphi )%
].  \label{tw28}
\end{equation}

These functions satisfy the orthogonality condition

\begin{equation}
\int \int Y^{*}(l,m^{(\widehat{\mathbf{a}})};\theta ,\varphi )Y(l,m^{\prime (%
\widehat{\mathbf{a}})};\theta ,\varphi )\sin \theta d\theta d\varphi =\delta
_{mm^{\prime }}.  \label{tw28a}
\end{equation}

\subsection{Probabilities}

The probabilities corresponding to the generalized spherical harmonics are
given by $P(1,m^{(\widehat{\mathbf{a}})};\theta ,\varphi )=\left| Y(1,m^{(%
\widehat{\mathbf{a}})};\theta ,\varphi )\right| ^2.$ They are

\begin{eqnarray}
P(1,1^{(\widehat{\mathbf{a}})};\theta ,\varphi ) &=&\frac 3{8\pi }[\cos
^2\theta ^{\prime }\sin ^2\theta \cos ^2(\varphi ^{\prime }-\varphi ) 
\nonumber \\
&&+\sin ^2\theta ^{\prime }\cos ^2\theta +\sin ^2\theta \sin ^2(\varphi
^{\prime }-\varphi )  \nonumber \\
&&+\frac 12\sin 2\theta \sin 2\theta ^{\prime }\cos (\varphi ^{\prime
}-\varphi )],  \label{fo46}
\end{eqnarray}

\begin{eqnarray}
P(1,0^{(\widehat{\mathbf{a}})};\theta ,\varphi ) &=&\frac 3{4\pi }[\cos
^2\theta ^{\prime }\cos ^2\theta +\sin ^2\theta \sin ^2\theta ^{\prime }\cos
^2(\varphi ^{\prime }-\varphi )  \nonumber \\
&&+\frac 12\sin 2\theta \sin 2\theta ^{\prime }\cos (\varphi ^{\prime
}-\varphi )]  \label{fo47}
\end{eqnarray}
and

\begin{eqnarray}
P(1,(-1)^{(\widehat{\mathbf{a}})};\theta ,\varphi ) &=&\frac 3{8\pi }[\cos
^2\theta ^{\prime }\sin ^2\theta \cos ^2(\varphi ^{\prime }-\varphi ) 
\nonumber \\
&&\ +\sin ^2\theta ^{\prime }\cos ^2\theta +\sin ^2\theta \sin ^2(\varphi
^{\prime }-\varphi )  \nonumber \\
&&\ -\frac 12\sin 2\theta \sin 2\theta ^{\prime }\cos (\varphi ^{\prime
}-\varphi )].  \label{fo48}
\end{eqnarray}

If $P(1,m^{(\widehat{\mathbf{a}})};\theta ,\varphi )$ is integrated over $%
(\theta ,\varphi ),$ the result is unity:

\begin{equation}
\iint P(1,m^{(\widehat{\mathbf{a}})};\theta ,\varphi )d\Omega =1.
\label{fo49}
\end{equation}

We observe that if we set $\theta ^{\prime }=\varphi ^{\prime }=0,$ so that $%
\widehat{\mathbf{a}}=\widehat{\mathbf{k}}$, we obtain the standard results 
\begin{equation}
P(1,1^{(\widehat{\mathbf{k}})};\theta ,\varphi )=\left| Y_1^1(\theta
,\varphi )\right| ^2=\frac 3{8\pi }\sin ^2\theta ,  \label{fi50}
\end{equation}

\begin{equation}
P(1,0^{(\widehat{\mathbf{k}})};\theta ,\varphi )=\left| Y_0^1(\theta
,\varphi )\right| ^2=\frac 3{4\pi }\cos ^2\theta  \label{fi51}
\end{equation}
and 
\begin{equation}
P(1,(-1)^{(\widehat{\mathbf{k}})};\theta ,\varphi )=\left| Y_{-1}^1(\theta
,\varphi )\right| ^2=\frac 3{8\pi }\sin ^2\theta .  \label{fi52}
\end{equation}

\subsection{Generalized Operators}

We expect that the generalized spherical harmonics are solutions of an
eigenvalue equation. In order to deduce this equation, we look more
carefully at the general features of differential eigenvalue equations.
Consider the time-independent Schr\"odinger equation; this is obtained by
forming an eigenvalue equation for the Hamiltonian operator $H(\mathbf{r}).$
Now, the Hamiltonian is expressed in terms of the position coordinate $%
\mathbf{r,}$ which also acts as the eigenfunction characterising the final
state. Similarly, in the Legendr\'e equation, the operator for the square of
the total angular momentum, $L^2,$ is expressed in terms of the angles $%
(\theta ,\varphi ),$ which also define the eigenvalue corresponding to the
final state. In both these cases, the variables which form the arguments of
the eigenfunction define the final eigenvalue pertaining to the measurement
which this eigenfunction is the probability amplitude for. Furthermore, we
observe that the variables corresponding to the final eigenvalue are the
ones in terms of which the derivatives in the differential operator are
defined. We therefore conclude that this is a general rule, and will use it
in the construction of the operator for the generalized spherical harmonics.

With these observations we are able to deduce the required operators. In the
primary coordinate system S defined by the unit vectors $\widehat{\mathbf{i}}
$, $\widehat{\mathbf{j}}$ and $\widehat{\mathbf{k}}$, the various angular
momentum operators are 
\begin{equation}
L_x=-i\hbar [-\sin \varphi \frac \partial {\partial \theta }-\cot \theta
\cos \varphi \frac \partial {\partial \varphi }],  \label{th30}
\end{equation}

\begin{equation}
L_y=-i\hbar [\cos \varphi \frac \partial {\partial \theta }-\cot \theta \sin
\varphi \frac \partial {\partial \varphi }],  \label{th31}
\end{equation}

\begin{equation}
L_z=-i\hbar \frac \partial {\partial \varphi }  \label{th32}
\end{equation}
and 
\begin{equation}
\mathbf{L}^2=-\hbar ^2[\frac 1{\sin \theta }\frac \partial {\partial \theta
}(\sin \theta \frac \partial {\partial \theta })+\frac 1{\sin ^2\theta }%
\frac{\partial ^2}{\partial \varphi ^2}].  \label{th33}
\end{equation}

In a new orthogonal coordinate system S' defined by the unit vectors $%
\widehat{\mathbf{u}}$, $\widehat{\mathbf{v}}$ and $\widehat{\mathbf{w}}$,
corresponding to the $x^{\prime }$, $y^{\prime }$ and $z^{\prime }$
directions, the operators are

\begin{equation}
L_x^{^{\prime }}=\widehat{\mathbf{u}}\cdot \mathbf{L,\;\;}L_y^{^{\prime }}=%
\widehat{\mathbf{v}}\cdot \mathbf{L}\text{ and }L_z^{^{\prime }}=\widehat{%
\mathbf{w}}\cdot \mathbf{L.}  \label{th34}
\end{equation}

Let the polar angles of the unit vectors $\widehat{\mathbf{u}}$, $\widehat{%
\mathbf{v}}$ and $\widehat{\mathbf{w}}$ be $(\theta _{\widehat{\mathbf{u}}%
},\varphi _{\widehat{\mathbf{u}}})$, $(\theta _{\widehat{\mathbf{v}}%
},\varphi _{\widehat{\mathbf{v}}})$ and $(\theta _{\widehat{\mathbf{w}}%
},\varphi _{\widehat{\mathbf{w}}})$ respectively. These form an orthogonal
set of axes, and so 
\begin{equation}
\widehat{\mathbf{u}}\cdot \widehat{\mathbf{v}}=\widehat{\mathbf{u}}\cdot 
\widehat{\mathbf{w}}=\widehat{\mathbf{v}}\cdot \widehat{\mathbf{w}}=0.
\label{th35}
\end{equation}
In addition, they form a right-handed coordinate system, so that 
\begin{equation}
\widehat{\mathbf{u}}\times \widehat{\mathbf{v}}=\widehat{\mathbf{w}},\;\;%
\widehat{\mathbf{v}}\times \widehat{\mathbf{w}}=\widehat{\mathbf{u}}\text{ \
and \ \ }\widehat{\mathbf{w}}\times \widehat{\mathbf{u}}=\widehat{\mathbf{v}}%
.  \label{th36}
\end{equation}
We can define all the unit vectors in terms of one set of polar angles $%
(\theta ^{\prime },\varphi ^{\prime })=(\theta _{\widehat{\mathbf{w}}%
},\varphi _{\widehat{\mathbf{w}}}),$ with the angles $(\theta _{\widehat{%
\mathbf{u}}},\varphi _{\widehat{\mathbf{u}}})$ and $(\theta _{\widehat{%
\mathbf{v}}},\varphi _{\widehat{\mathbf{v}}})$ defined in such a way as to
ensure that the conditions Eqs. (\ref{th35}) and (\ref{th36}) are satisfied.
This means that the unit vector $\widehat{\mathbf{w}}$ is to be identified
with the unit vector $\widehat{\mathbf{a}}$. One choice for the vectors $%
\widehat{\mathbf{u}}$ and $\widehat{\mathbf{v}}$ is 
\begin{equation}
\theta _u=\theta ^{\prime }-\pi /2,\;\;\varphi _u=\varphi ^{\prime
};\;\;\theta _v=\pi /2,\;\;\varphi _v=\varphi ^{\prime }-\pi /2.
\label{th37}
\end{equation}
Thus, while 
\begin{equation}
\widehat{\mathbf{w}}=(\sin \theta ^{\prime }\cos \varphi ^{\prime },\sin
\theta ^{\prime }\sin \varphi ^{\prime },\cos \theta ^{\prime }),
\label{th38}
\end{equation}
we have

\begin{equation}
\widehat{\mathbf{u}}=(-\cos \theta ^{\prime }\cos \varphi ^{\prime },-\cos
\theta ^{\prime }\sin \varphi ^{\prime },\sin \theta ^{\prime })
\label{th39}
\end{equation}
and 
\begin{equation}
\widehat{\mathbf{v}}=(\sin \varphi ^{\prime },-\cos \varphi ^{\prime },0).\;
\label{fo40}
\end{equation}
Using Eqs. (\ref{th30})- (\ref{th32})\ and Eqs. (\ref{th34}), we find that
the new operators are

\begin{equation}
L_x^{^{\prime }}=-i\hbar \left( \cos \theta ^{\prime }\sin (\varphi -\varphi
^{\prime })\frac \partial {\partial \theta }+[\cos \theta ^{\prime }\cot
\theta \cos (\varphi -\varphi ^{\prime })+\sin \theta ^{\prime }]\frac
\partial {\partial \varphi }\right) ,  \label{fo41}
\end{equation}
\begin{equation}
L_y^{^{\prime }}=i\hbar \left( \cos (\varphi -\varphi ^{\prime })\frac
\partial {\partial \theta }-\cot \theta \sin (\varphi -\varphi ^{\prime
})\frac \partial {\partial \varphi }\right)  \label{fo42}
\end{equation}
and 
\begin{equation}
L_z^{^{\prime }}=i\hbar \left( \sin \theta ^{\prime }\sin (\varphi -\varphi
^{\prime })\frac \partial {\partial \theta }+[\sin \theta ^{\prime }\cot
\theta \cos (\varphi -\varphi ^{\prime })-\cos \theta ^{\prime }]\frac
\partial {\partial \varphi }\right) .  \label{fo43}
\end{equation}
The square of the total angular momentum remains the same, of course. If our
reasoning is correct, the generalized spherical harmonics should be
eigenfunctions of $L_z^{^{\prime }}$ and $\mathbf{L}^2$, Eqs. (\ref{fo43})
and Eq. (\ref{th33}), respectively. We find that this is indeed so:

\begin{equation}
\mathbf{L}^2Y(l,m^{(\widehat{\mathbf{a}})};\theta ,\varphi )=l(l+1)\hbar
^2Y(l,m^{(\widehat{\mathbf{a}})};\theta ,\varphi )  \label{fo44}
\end{equation}
and 
\begin{equation}
L_z^{^{\prime }}Y(l,m^{(\widehat{\mathbf{a}})};\theta ,\varphi )=m^{(%
\widehat{\mathbf{a}})}\hbar Y(l,m^{(\widehat{\mathbf{a}})};\theta ,\varphi ).
\label{fo45}
\end{equation}

\subsection{Eigenvalue Equations for $L_x^{^{\prime }}$ and $L_y^{^{\prime
}} $}

There is a simple prescription for obtaining $L_x^{^{\prime }}$ and $%
L_y^{^{\prime }}$ when the generalized expression for $L_z^{^{\prime }}$ is
known[2]. To obtain $L_x^{^{\prime }}$ from $L_z^{^{\prime }}$ we simply
make the transformation $\theta ^{\prime }\rightarrow \theta ^{\prime }-\pi
/2,$ leaving $\varphi ^{\prime }$ unchanged. To obtain $L_y^{^{\prime }}$ we
set $\theta ^{\prime }=\pi /2\;$ and $\;\varphi ^{\prime }\rightarrow
\varphi ^{\prime }-\pi /2.$ These transformations also convert the
eigenfunctions of $L_z^{^{\prime }}$ into the eigenfunctions of $%
L_x^{^{\prime }}$ or $L_y^{^{\prime }}$, as the case may be.$_{\text{ }}$ We
consider first the case of $L_x^{^{\prime }};$ its eigenvalue equation is

\begin{equation}
L_x^{^{\prime }}Y(l,m^{(\widehat{\mathbf{u}})};\theta ,\varphi )=m^{(%
\widehat{\mathbf{u}})}\hbar Y(l,m^{(\widehat{\mathbf{u}})};\theta ,\varphi ).
\label{fi53}
\end{equation}
The eigenfunctions of the operator Eq. (\ref{fo41}) are obtained from the
generalized spherical harmonics Eqs. (\ref{tw26})-(\ref{tw28}) by applying
to these functions the transformation $\theta ^{\prime }\rightarrow \theta
^{\prime }-\pi /2$. The eigenfunctions of $L_x^{^{\prime }}$ are found to be

\begin{equation}
Y(1,1^{(\widehat{\mathbf{u}})};\theta ,\varphi )=-\sqrt{\frac 3{8\pi }}[\cos
\theta ^{\prime }\cos \theta +\sin \theta [\sin \theta ^{\prime }\cos
(\varphi ^{\prime }-\varphi )-i\sin \theta \sin (\varphi ^{\prime }-\varphi
)]],  \label{fi54}
\end{equation}

\begin{equation}
Y(1,0^{(\widehat{\mathbf{u}})};\theta ,\varphi )=\sqrt{\frac 3{4\pi }}[\sin
\theta ^{\prime }\cos \theta -\cos \theta ^{\prime }\sin \theta \cos
(\varphi ^{\prime }-\varphi )]  \label{fi55}
\end{equation}
and

\begin{equation}
Y(1,(-1)^{(\widehat{\mathbf{u}})};\theta ,\varphi )=\sqrt{\frac 3{8\pi }}[%
\cos \theta ^{\prime }\cos \theta +\sin \theta [\sin \theta ^{\prime }\cos
(\varphi ^{\prime }-\varphi )+i\sin \theta \sin (\varphi ^{\prime }-\varphi
)]].  \label{fi56}
\end{equation}

The interpretation of these eigenfunctions is evident: they are probability
amplitudes for measurements of the angular position $(\theta ,\varphi )$\ if
the system is initially in the state of angular momentum projection $m\hbar $
in the direction $\widehat{\mathbf{u}}.$

We next shift attention to $L_y^{^{\prime }}$, whose eigenvalue equation is 
\begin{equation}
L_x^{^{\prime }}Y(1,m^{(\widehat{\mathbf{v}})};\theta ,\varphi )=m^{(%
\widehat{\mathbf{v}})}\hbar Y(1,m^{(\widehat{\mathbf{v}})};\theta ,\varphi ).
\label{fi57}
\end{equation}
The transformations $\theta ^{\prime }=\pi /2\;$ and $\;\varphi ^{\prime
}\rightarrow \varphi ^{\prime }-\pi /2$ applied to the expression for $%
L_z^{^{\prime }}$ transform this operator to $L_y^{^{\prime }}.$ The same
transformations applied to the eigenfunctions of $L_z^{^{\prime }}$ yield
from them the eigenfunctions of $L_y^{^{\prime }}.$ The eigenvalue equation
is then found to have the eigenfunctions

\begin{equation}
Y(1,1^{(\widehat{\mathbf{v}})};\theta ,\varphi )=\sqrt{\frac 3{8\pi }}[\cos
\theta +i\sin \theta \cos (\varphi ^{\prime }-\varphi )],  \label{fi58}
\end{equation}

\begin{equation}
Y(1,0^{(\widehat{\mathbf{v}})};\theta ,\varphi )=\sqrt{\frac 3{4\pi }}\sin
\theta \sin (\varphi ^{\prime }-\varphi )]  \label{fi59}
\end{equation}
and

\begin{equation}
Y(1,(-1)^{(\widehat{\mathbf{v}})};\theta ,\varphi )=-\sqrt{\frac 3{8\pi }}[%
\cos \theta +i\sin \theta \cos (\varphi -\varphi ^{\prime })].  \label{si60}
\end{equation}

The eigenfunctions $Y(1,m^{(\widehat{\mathbf{v}})};\theta ,\varphi )$ of $%
L_y^{^{\prime }}$ are probability amplitudes for measurements of the angular
position of the system if its initial state corresponds to the angular
momentum projection $m\hbar $ in the direction $\widehat{\mathbf{v}}$.

\section{Conclusion}

In this paper we have deduced generalized spherical harmonics and the
eigenvalue equation they satisfy. It is striking that after postulating the
generalized spherical harmonics, we were able to deduce the eigenvalue
equation by plausibility arguments. Furthermore, the generalized spherical
harmonics for the various components of the angular momentum have a
reasonable and obvious interpretation. The next step is to extend these
ideas to the case $l=2,$ and indeed higher values of $l$. This work is in
progress.

\subsection{References}

1. Mweene H. V., ''Derivation of Spin Vectors and Operators From First
Principles'' , quant-ph/9905012.

2. Mweene H. V., ''Generalised Spin-1/2 Operators And Their Eigenvectors'',
quant-ph/9906002.

3. Mweene H. V., ''Vectors and Operators For Spin 1 Derived From First
Principles'', quant-ph/9906043.

4. Mweene H. V., ''Alternative Forms of Generalized Vectors and Operators
for Spin 1/2'', quant-ph/9907031.

5. Mweene H. V., ''Spin Description and Calculations in the Lande'
Interpretation of Quantum Mechanics'', quant-ph/9907033.

6. Mweene H. V., ''New Treatment of Systems of Compounded Angular
Momentum'', quant-ph/9907082.

7. Mweene H. V., ''Derivation of Standard Treatment of Spin Addition From
Probability Amplitudes'', quant-ph/0003056.

8. Land\'e A., ''From Dualism To Unity in Quantum Physics'', Cambridge
University Press, 1960.

9. Land\'e A., ''New Foundations of Quantum Mechanics'', Cambridge
University Press, 1965.

10. Land\'e A., ''Foundations of Quantum Theory,'' Yale University Press,
1955.

11. Land\'e A., ''Quantum Mechanics in a New Key,'' Exposition Press, 1973.

\end{document}